\documentclass[twoside,twocolumn,british,showpacs,preprintnumbers,amsmath,amssymb,groupedaddress,nofootinbib]{revtex4-1}
\usepackage[utf8]{inputenc}
\usepackage[a4paper]{geometry}
\usepackage{amsmath}
\usepackage{amssymb}
\usepackage{graphicx}
\usepackage{esint}

\makeatletter
\usepackage{babel}

\usepackage{psfrag}
\usepackage{subfigure}
\usepackage{amscd}
\usepackage{amsfonts}
\usepackage{here}

\let\textquotedbl="

\usepackage{babel}

\makeatother

\usepackage{babel}
\begin{document}

\title{Role of Helium-Hydrogen ratio on energetic interchange mode behaviour
and its effect on ion temperature and micro-turbulence in LHD }

\author{C.A. Michael$^{1,2*}$, K. Tanaka$^{2,3}$, T. Akiyama$^{2,4}$,
T. Ozaki$^{2}$, M. Osakabe$^{2,4}$, S. Sakakibara$^{2}$, H. Yamaguchi$^{2}$,
S. Murakami$^{5}$, M. Yokoyama$^{2,4}$, M. Shoji$^{2}$, L.N. Vyacheslavov$^{6}$
and LHD experimental group$^{2}$ }

\address{$^{1}$Australian National University, Canberra, A.C.T. 2601, Australia}

\address{$^{2}$ National Institute for Fusion Science, National Institutes
of Natural Science, 322-6 Oroshi-cho, Toki, Japan 509-5292}

\address{$^{3}$ Department of Advanced Energy and Engineering, Kyushu University,
Kasuga, Fukuoka 816-8580, Japan}

\address{$^{4}$ SOKENDAI (The Graduate University for Advanced Studies),
Toki, Gifu 509-5292, Japan}

\address{$^{5}$ Department of Nuclear Engineering, Kyoto University, Kyoto
615-8540 Japan}

\address{$^{6}$ Budker Institute of Nuclear Physics SB RAS, 630090, Novosibirsk,
Russian Federation }

\date{\today}
\begin{abstract}
In the Large helical device, a change of energetic particle mode is
observed as He concentration is varied in ion-ITB type experiments,
having constant electron density and input heating power but with
a clear increase of central ion temperature in He rich discharges.
This activity consists of bursty, but damped energetic interchange
modes (EICs, X Du et al., Phys. Rev. Lett. 114 p.155003 (2015)), whose
occurrence rate is dramatically lower in the He-rich discharges. Mechanisms
are discussed for the changes in drive and damping of the modes with
He concentration. These EIC bursts consist of marked changes in the
radial electric field, which is derived from the phase velocity of
turbulence measured with the 2D phase contrast imaging (PCI) system.
Similar bursts are detected in edge fast ion diagnostics. Ion thermal
transport by gyro-Bohm scaling is recognised as a contribution to
the change in ion temperature, though fast ion losses by these EIC
modes may also contribute to the ion temperature dependence on He
concentration, most particularly controlling the height of an ``edge-pedestal''
in the $T_{i}$ profile. The steady-state level of fast ions is shown
to be larger in Helium rich discharges on the basis of a compact neutral
particle analyser (CNPA), and the fast-ion component of the diamagnetic
stored energy. These events also have an influence on turbulence and
transport. The large velocity shear induced produced during these
events transiently improves confinement and suppresses turbulence,
and has a larger net effect when bursts are more frequent in Hydrogen
discharges. This exactly offsets the increased gyro-Bohm related turbulence
drive in Hydrogen which results in the same time-averaged turbulence
level in Hydrogen as in Helium. 
\end{abstract}
\maketitle

\section{Introduction}

\renewcommand{\thefootnote}{\fnsymbol{footnote}} \footnote[0]{* C.A. Michael was a visiting associate professor from June to July 2016.}The
differences of confinement of H and D plasmas, the so-called ``isotope
effect'' in Tokamaks, reversed-field pinches (RFPs) and Stellarators/Heliotrons
has long been a largely theoretically unresolved problem. In Tokamaks
and RFPs, Deuterium is almost always noted to have better confinement
than Hydrogen \cite{isotope_jt60,lorenzini}. However, the main scaling
law for confinement, so-called Bohm or Gyro-Bohm scaling laws predict
that Deuterium should have worse confinement due to a larger gyro-radius
\cite{asdex_isotope}. One mechanism considered to relate to this
is the change in damping rate of geodesic acoustic modes (GAMS). \cite{gurchenko_gams_isotope}.
On the other hand, in Stellarators, early experiments in W7-AS \cite{isotope_stellarator_w7as}
suggested that there was only a very modest (20\%) increase in electron
tempearture in Deuterium, and a database confinement study found virtually
no difference in the ISS-normalized energy confinement time between
Deuterium and Hydrogen in W7-AS, Heliotron-E and the ATF torastron
\cite{yamada_fst}. Furthermore, improved particle transport in Deuterium
has been reported in CHS \cite{isotope_tanaka_chs} and Heliotron-J
\cite{ohtani_heliotron_j}. Recently there has been an impetus to
study these effects on the Large Helical Device (LHD), which has motivated
the planned Deuterium campaign. Prior to starting this, a series of
experiments were conducted in order to make a comparison of H and
He discharges, which may give insight into both mass (A) and charge
(Z) dependencies of transport (in particular the gyro-Bohm dependence
which goes as $\sqrt{A}/Z^{2}$), heating and instabilities. This
has been the focus of several recent investigations of heat transport
in the Large helical device \cite{tanaka_nf2017}, examining the role
ion density (which reduces relative to electron density as the mass
increases), absorbed heating power and transport, both neoclassical
and turbulent. It was found that the collisionality dependence was
much stronger than the ion species effects,  such that at high collisionality,
the ion-species had little distinguishable effect \cite{tanaka_nf2017}.
However, at low collisionality, characteristic of the ion-ITB type
mode \cite{nagaoka_ti_scenario}, the ion temperature increases with
Helium concentration, by an amount significantly exceeding the gyro-Bohm
scaling factor, particularly towards the edge \cite{tanaka_nf2017}.
It was shown \cite{nagaoka} that this could not be explained by change
of heating power, as this was mostly independent of He concentration,
despite the ion density going down. It was shown in \cite{nunami_cmp_h_he}
that the change of $T_{e}/T_{i}$ could contribute to this increase
in confinement. The properties of turbulence in these discharges were
analysed in \cite{tanaka_nf2017,nunami_cmp_h_he} and it was shown
that the time-averaged density fluctuation profile, measured using
phase-contrast imaging \cite{tanaka_pci} was the same in both H and
He discharges which is difficult to reconcile with the confinement
improvement. In addition to the properties of heating and confinement,
there is the possibility that energetic particle instabilities, driven
by strong NBI heating, can also be affected by change of species.

Energetic particle instabilities, such as toroidal and beta-induced
Alfven eigenmodes (TAEs and BAEs) and fishbones are known to significantly
degrade the population of NBI or ICRF-heated ions in Tokamaks \cite{heibrink_review}.
Fast-electron driven instabilities have also been identified, including
e-TAEs \cite{r_etae}, e-BAEs \cite{r_ebae} and e-fishbones \cite{r_efishbone}.
Additionally, according to simulation, micro-turbulence driven by
spatial gradients of bulk plasma properties (ion temperature) can
also lead to a loss of energetic particles \cite{jenko_turb_fastion}.
In fact, it is essential for heat transport analysis to validate predictions
of plasma heating using neutral beams and ICRH based on fast ion measurements
using well-established absolutely calibrated diagnostics such as fission
chambers \cite{fi_transport_mast}. However, in Stellarators and Heliotrons,
such systems are less mature in part owing the complexity in modelling
and measurement of fast ions. On LHD, several types of energetic particle
instabilities have been identified, for example TAEs, GAEs \cite{toi2011},
EAEs, fishbones \cite{toi2000}, and E-GAMS \cite{egam} which are
also in the range of frequency of BAEs. Recently, however, a new
type of fast ion instability was discovered, the energetic particle
interchange mode (EIC) which results from a resonance of the precession
frequency (which is around 7kHz for the m=1,n=1 mode) of perpendicularly-injected
fast ions (E=34keV) with the resistive interchange mode which is located
at around r/a=0.85, near the $\iota=1/1$ rational surface \cite{eic_nf,eic_prl,eic_theor,nishimura_sim_eic_2016}.
These tend to occur in low density plasmas ($n_{e}<1.5\times10^{19}\mathrm{m}^{-3}$)
and are common at high heating power of perpendicular neutral beam
injection \cite{eic_nf}. These modes tend to be bursting in time
(repeating at various intervals in the order of \textasciitilde{}0.1s),
lasting around 2-3ms, and exhibit chirp-down behaviour associated
with a non-linear wave/particle interaction. Associated with fast
ion redistribution, the plasma potential and toroidal rotation were
shown to change dramatically within each burst.

In this paper, it is shown that EIC modes exists in the ion-ITB discharges
used for comparing ion species effects, but are ``damped'' in that
only a single oscillation spike exists (at around 7-8kHz), and that
the perturbation is localized both poloidally and toroidally. It was
shown in \cite{eic_nf} that this behaviour tends to occur near the
marginal boundary of existence of these modes (at higher density or
marginal power). These burst rate of these modes decreases in He-rich
discharges where $T_{i}$ is higher, which may give clues as to the
stability threshold for this mode. Also in these discharges, a steady
30kHz mode is found, with m=4,n=1 mode structure. Although it seems
like the 4th harmonic of an EIC mode, it does not fit the resonance
condition for deeply trapped perpendicularly-injected fast ions as
the mode propagates in the electron diamagnetic direction, whilst
the trapped ions precess in the ion-diamangetic direction.

It is vital also to consider how the fast ion population varies with
He concentration, which could be either directly as a consequence
of the CX neutral losses of energetic ions, or even the change of
confinement associated with the gyro-Bohm scaling. Consideration is
also given to the direct losses associated with the MHD modes. Analysis
of the fast ion population is carried out by comparing CNPA signals,
as well as the diamagnetic stored energy. The main tool for investigation
is the 2D phase-contrast imaging system \cite{tanaka_pci} which was
analysed with very narrow time window (\textless{}0.1ms) and reveals
dynamics in terms of large changes of the phase velocity of turbulence,
which, given it is dominated by $E\times B$ drift, indicates the
$E_{r}$ arising due to the fast ion losses and/or redistribution.
Turbulence is suppressed during the bursts as reported in previous
work \cite{tanaka_nf2017,eic_prl}, and an analysis is carried out
to link these observations with the confinement improvement in He.

This paper is organised as follows. In Sec II, the reference discharges
are presented including the change of ion species concentration and
temperature, together with previous studies on transport, turbulence
measurements gyro-kinetic simulations. In Sec. III, the observations
of the fast-ion MHD events are presented, focusing on changes to the
phase velocity of PCI measurements, and results are presented on how
these change with the Helium concentration, and the frequency and
mode numbers are analysed. Also, evidence is presented to demonstrate
the loss of fast ions during these events. A detailed analysis of
the temporal modulation of turbulence amplitude and electron temperature
by the MHD bursts is carried out, focusing on transport improvement.
In Sec IV, we examine how the fast ion distribution, most particularly,
the component driven by perpendicular NBI, changes with Helium concentration,
based on NPA and diamagnetic loop signals. In Sec V, we discuss physics
issues raised by these observations, looking at background neutrals,
the fast ion distribution function, drive damping sources for this
mode, and we unify the evidence for the EIC's influence on ion heating.
Finally, in Sec. VI conclusions are presented. 

In this paper, we utilize the cross-covariance $R_{xy}(L)$ for a
given lag L, between the two variables x and y. We define the conventional
correlation $\gamma_{xy}(L)$, the correlation $\gamma_{xy}^{chg}(L)$
which quantifies how a change in $x$ occurs after averaging over
a series of bursts in variable y (for which we use phase velocity
at $\rho=0.75$), and $\gamma_{xy}^{rel}(L)$, the normalization which
represents the relative change in x : 

\begin{equation}
\gamma_{xy}(L)=\frac{R_{xy}(L)}{\sqrt{R_{xx}(0)R_{yy}(0)}}\label{eq:gamma}
\end{equation}

\begin{equation}
\gamma_{xy}^{chg}(L)=\frac{R_{xy}(L)}{\sqrt{R_{yy}(0)}}\label{eq:gammachg}
\end{equation}

\begin{equation}
\gamma_{xy}^{rel}(L)=\frac{R_{xy}(L)}{\overline{x}\sqrt{R_{yy}(0)}}.\label{eq:gammarel}
\end{equation}

\emph{}

\section{Ion-ITB comparison discharges\label{sec:Ion-ITB-comparison-discharges}}

\begin{figure}
\includegraphics[width=8cm]{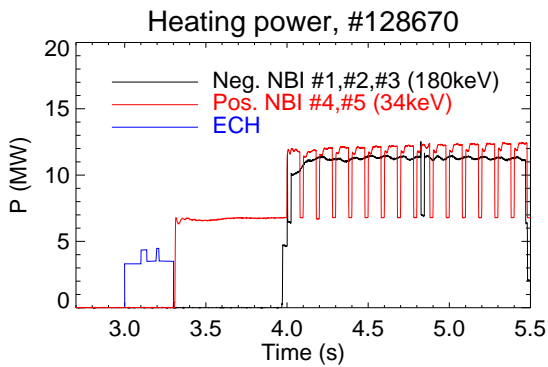}

\caption{Heating power waveform, which is the same in all H/He comparison discharges.\label{fig:Discharge-scenario}}
\end{figure}

The targeted scenario for these comparison discharges was one with
an ion internal transport barrier (ITB) \cite{tanaka_highti,ita_itb},
where the strong ion heating results in high ion temperature, and
the generation of ion temperature gradient (ITG) modes at mid radius.
However, the transport ``barrier'' is considered to be due to a
reduction of the gyro-Bohm normalized ion thermal conductivity at
mid radius \cite{ita_itb}. Subsequent to these discharges, the ion
heating power was upgraded to provide further heating from an additional
low energy positive-ion based neutral beam injection system. The heating
waveform for this study is presented in Fig. (\ref{fig:Discharge-scenario}).
There are five Hydrogen neutral beam injection systems on LHD, with
three of them injecting in the tangential direction to the field,
and two of them injecting perpendicularly, with one of these beams
being modulated at 10Hz (80ms on, 20ms off) for CXRS background measurement
to obtain ion temperatures. For these shots, the perpendicular beams
have an energy of 39-45keV whilst the tangential beams have an energy
of 164-178keV (varying between each beam source but not between discharges)\cite{lhd_nbi}.
These high energy beams are well above the ``critical energy'' and
deposit most of their power into electrons, whilst the low energy
beams deposit most of their absorbed power into ions. For the perpendicular
beams, there can be a relatively large shine through fraction, particularly
at lower densities, because of the small path length through the plasma,
and the tendency for fast ions with high pitch angle to be subject
to prompt orbit loss. It is estimated that only 30\% of the perpendicularly-injected
neutral beam power is absorbed by ions.

Four discharges are highlighted where the neutral Helium concentration
in the fuelling gas is varied, whilst the electron density is kept
constant. The passive neutral spectral lines of He ($\lambda=587\textrm{nm}$)
and H ($\lambda=656$nm) \cite{goto_spectroscopy}, which relate to
the influx near the edge of the plasma, are used to measure the relative
concentration of Helium in the discharge defined as $c_{He}=n_{He}/(n_{H}+n_{He}),$with
$c_{H}=1-c_{He}$ \cite{ida_influx}. In fact, the ratio of H:He has
been shown the rather flat over the entire profile using CX spectroscopy
\cite{ida_hheprofile}. The values of $c_{He}$ vary from $20\%$
to $70\%$, as neither Hydrogen or Helium could be completely pumped
out of the vessel and walls. In these discharges, the line-averaged
density exhibits different temporal dependence on account of the differing
gas fuelling, however, feedback control brings the central line-averaged
electron density to the same value at t=4.7s, and spatial profiles
of electron density are also very well-matched. Further details of
these discharges is discussed in Ref. \cite{nagaoka}. Whilst the
electron density is the same in all four discharges, the total density
of Helium and Hydrogen ions are derivable from quasi-neutrality:

\begin{align}
f_{He} & =n_{He}/n_{e}=(1-c_{H})/(2-c_{H})\label{eq:fhe}\\
f_{H} & =n_{H}/n_{e}=2c_{H}/(2-c_{H})\label{eq:fh}
\end{align}

\begin{figure}
\includegraphics[width=8cm]{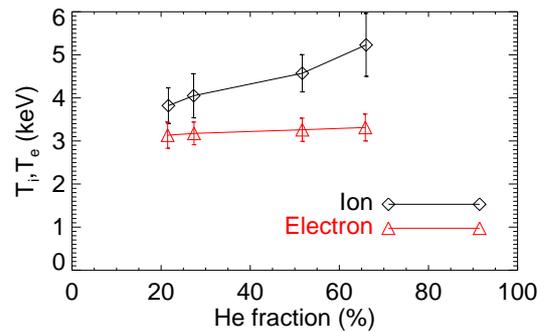}

\caption{Dependence of electron and ion temperature (averaged over the core
region $\rho<0.3,$ with variations conveyed in the error bar) on
Helium concentration. \label{fig:Dependence-of-electron}}
\end{figure}

From these equations, the total ion density fraction $f_{H}+f_{HE}$
decreases down to 1/2 with increase in He concentration, and is an
important consideration in the analysis, particularly of the heating
power per particle, as well as the contribution to the diamagnetic
energy. The dependence of the core values of both $T_{i}$ and $T_{e}$
on He concentration is captured in Fig. (\ref{fig:Dependence-of-electron}).
Clearly, the electron temperature does not change, whilst the ion
temperature increases by $\approx30\%$ in He rich discharges. The
first reason that may come to mind to explain this dependence is the
reduction of the ion density in He rich discharges, leading to a higher
power deposited per particle. However, using the TASK3D and GNET codes
\cite{nagaoka}, it was shown that the reduction in ion density in
Helium is offset by an decrease in the critical energy $E_{\textrm{cr}}\sim(Z/A)^{2/3}$
(at which the ion and electron heating powers are equal), leading
to less net power being coupled to the ions in He plasma. 

A detailed analysis of the $T_{i}$ profile, given in \cite{tanaka_nf2017},
shows $L_{Ti}^{-1}$ is rather constant (and the same in He and H
discharges) in the core regions ($\rho<0.9$), and increases in an
``edge-pedestal'' region. However, the height of the edge pedestal
is larger in He discharges, leading by invariance of the core $L_{Ti}^{-1}$
to a higher core $T_{i}$. This shows that ion transport is rather
stiff across the core regions so that the increase in $T_{i}$ at
the edge pedestal would contribute to raising $T_{i}$ at the centre.
Thus, the physics of edge temperature (heating, neoclassical, anomalous
transport) are key to this dependence. As discussed, calculations
of the heating power cannot explain the temperature difference \cite{nagaoka}. 

Anomalous transport is expected to follow gyro-Bohm dependence $\sqrt{A}/Z^{2}.$
Thus, it is natural to expect that the temperature would be greater
by a factor of 2 in a Helium discharge for the same heating power
(though other parameters such as ion temperature are part of the gyro-Bohm
scaling). Transport analysis was carried out on these discharges in
\cite{tanaka_nf2017} to derive the net (power-balance) thermal diffusivity,
as well as the neoclassical values. Surprisingly, the neoclassical
contribution exceeded the net thermal diffusivity in the core, however
it was concluded that the neoclassical values were untrustworthy in
the presence of high beam power and strong external torque, and in
fact the measured (rotation-based) radial electric field followed
the electron root towards the core, contrary to the neoclassical predictions.
Whilst it is possible from these calculations that the neoclassical
contribution is non-negligible in these discharges, the presence of
a critical gradient in the profile is good evidence that anomalous
transport plays a significant role.

Considering the anomalous transport contribution, it is appropriate
to normalise the transport coefficient with the gyro-Bohm scaling
factor. It was shown in \cite{tanaka_nf2017} that even after allowing
for the gyro-Bohm factor, ion thermal transport in Helium discharge
\#128670 is lower than in Hydrogen discharge \#128717 by a factor
$\sim$2 times in the peripheral regions ($\rho>0.8$). Further detailed
quasi-linear gyro-kinetic analysis of the heat diffusivity in these
discharges showed that the increase of $T_{i}$ itself contributes
partially to further reducing the gyro-Bohm normalised thermal diffusivity
in He discharges, but is insufficient in peripheral regions to entirely
explain this improvement. Additionally, it was shown that the experimental
values of $L_{Ti}^{-1}$ in H and He discharges were slightly above
the critical gradient for ITG turbulence (which did not change between
H and He, consistent with the experimental results over the inner
radii), however, the linear growth rate above this threshold was lower
in the He-rich discharge. 

In summary, gyro-kinetic simulations qualitatively agree with the
trend of the heat diffusivity vs He concentration as it exhibits a
clear gyro-Bohm scaling. The calculated linear growth rate is lower
in the He-rich discharge whilst the ion thermal diffusivity $\chi_{i}=q/n\nabla T_{i}$
is also lower by by ratio of the temperature gradient, given that
the GNET \cite{gnet_icrh} shows the power per particle to be the
same in H and He discharges. On the other hand, the change of turbulence
properties between H and He discharges is more complicated. According
to quasi-linear theory, the fluctuation amplitude should be proportional
to the growth rate, unless some other mechanism governing the saturation
level changes such as the peak wave-vector or zonal flow activity.
Thus, one might initially expect the H discharge to have a higher
level of turbulence. However, comparing the turbulence properties
from the 2D-PCI system \cite{tanaka_pci}, using a long time interval
(4.70-4.78s), it was shown that the amplitude profile was virtually
identical between He and H rich discharges \cite{nagaoka}. This inconclusive
result has motivated a closer examination of 2D PCI signals, which
revealed energetic particle-driven MHD bursts, with signatures appearing
more clearly in PCI than any other diagnostic \cite{tanaka_nf2017}.
The burst rate is larger in the H rich discharge, and, given that
these bursts result in a temporary suppression of the turbulence,
may be the reason that the average level of turbulence does not increase
for H-rich discharges, despite having a higher linear growth rate.

\section{Bursting Fast-Ion MHD activity\label{sec:Bursting-Fast-Ion-MHD}}

\subsection{Changes in turbulence and radial electric field }

\begin{figure*}
\includegraphics{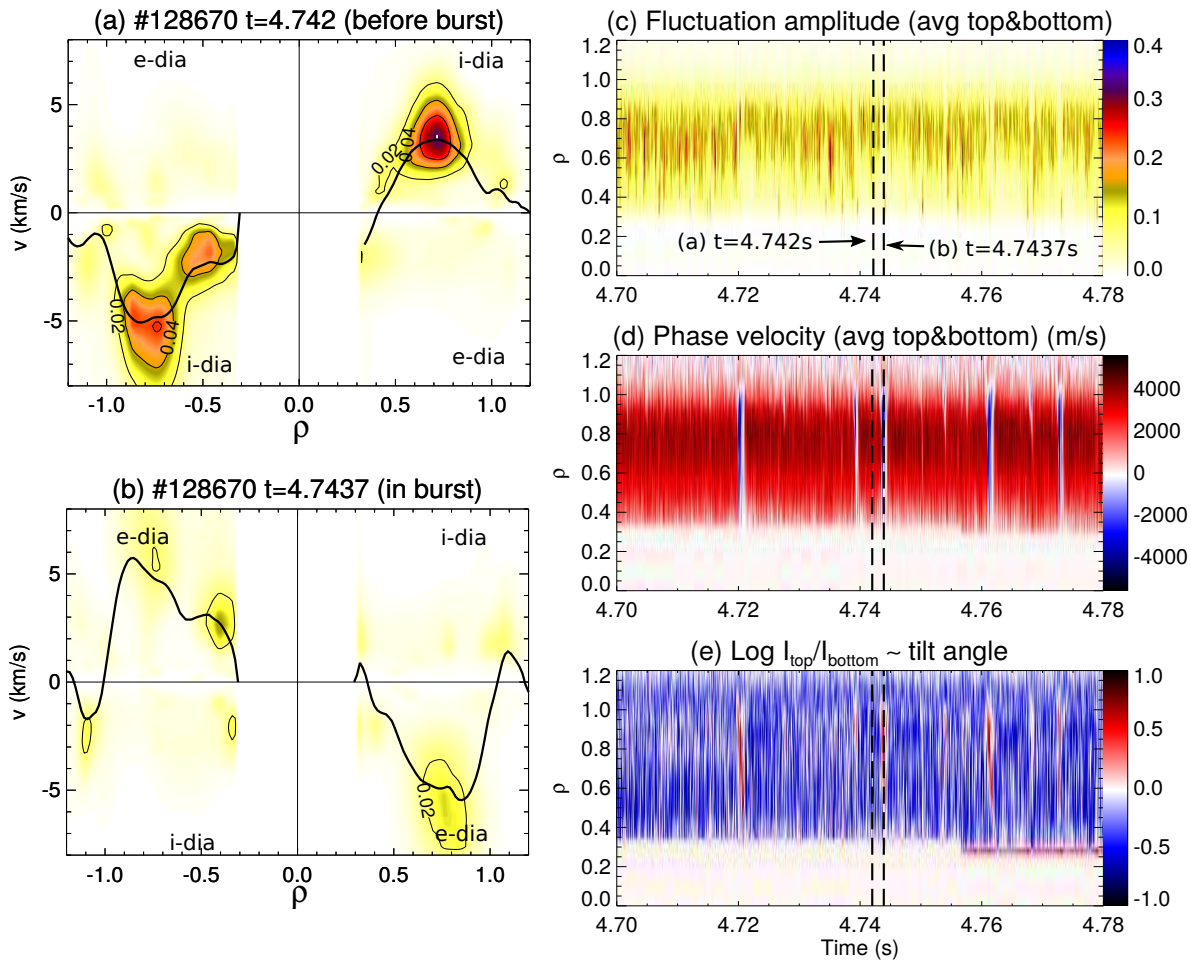}\caption{Turbulence spectral density $S(\rho,V)$, where $\rho$ is the normalized
radius (sign denoting which side of the mid-plane) and $V$ is the
phase velocity, at (a) t=4.742s and (b) 4.7437s in He dominant discharge
\#128670 (same linear colour scale and contour levels). The line indicates
the average phase velocity (projected perpendicular to vertical probing
beam), which is taken to greatly represent the $E\times B$ poloidal
rotation velocity of the plasma. These are compared with the time
histories of (c) fluctuation amplitude (linear colour scale) and (d)
fluctuation phase velocity spatial profiles, averaged over the bottom
and top halves of the profile, as well as (e) the amplitude asymmetry,
which relates to the eddy tilt angle.\label{fig:pciprofs_1_2}}
\end{figure*}

In these H/He comparison discharges, fast ion instabilities were identified
with multiple diagnostics, but primarily the initial tool was a 2D
phase contrast imaging system \cite{tanaka_pci}. This diagnostic
measures the line-integrated density fluctuations, and, through frequency
and k analysis, also enables the phase spatial profile (mapped to
flux coordinate $\rho$) of the fluctuation amplitude, as well as
the phase velocity profile $V$ (perpendicular to the line of sight)
to be obtained. An analysis of discharges in \#128717 (The H-rich
shot) was carried out considering very short time windows (0.1ms),
as shown in Fig. (\ref{fig:pciprofs_1_2}). Consequently, only turbulence
components with f\textgreater{}10kHz have been considered. In this
manner, 2D PCI can be used to deliver information about the time variation
of the radial electric field with arbitrarily high time resolution
(limited only by the bandwidth of the turbulence). The spectral density
of fluctuations $S(\rho,V)$ is plotted in Fig. (\ref{fig:pciprofs_1_2}a,b).
The gap at $-0.5<\rho<0.5$ is because the sight line does not penetrate
to the magnetic axis. Furthermore, the sign of $\rho$ denotes locations
above (positive) and below the mid-plane (negative). It can be sensitive
to the eddy tilt angle and so this can cause an apparent up/down asymmetry.
The average velocity at each location is calculated using the moment
over velocity, and is denoted by the solid line. Since the diagnostic
measures only the component of the velocity perpendicular to the vertical
probing beam, and since turbulence propagates in the poloidal direction
(including $E\times B$ and diamagnetic rotation components), the
component of measured phase velocity is high near the edge of the
plasma, reducing towards zero at the tangency radius, and inverting
on the other side. The phase velocity is dominated by $E\times B$
rotation, as has been shown in \cite{tanaka_nf2017}. The key interpretation
here is that turbulence phase velocity is dominated by $E\times B$
rotation and so this can be a monitor at high time resolution of changes
in electric field. 

The comparison between Figs (\ref{fig:pciprofs_1_2}a) and (b) shows
how the turbulence velocity spectra change spontaneously within a
short period of time. Initially, the phase velocity is in the ion
diamagnetic direction, which is typical for high $T_{i}$ discharges,
and then, it reverses spontaneously towards the electron diamagnetic
direction. It is suggested that fast ion losses are responsible for
the change of direction, and that this mode is very much like an energetic
interchange mode, on the basis of magnetic probe measurements in Sec.
(\ref{subsec:Magnetic-probe-signatures}) and fast ion loss measurements
presented in Sec. (\ref{subsec:Fast-ion-loss}). Whilst the poloidal
rotation was not available in this discharge, heavy-ion beam-probe
data indicates that in EIC events \cite{eic_nf} that the radial electric
field reverses during the burst. In addition, the symmetry of the
turbulence signal changes. This is most likely to be due to a change
in the eddy tilt angle, which results in constructive interference
on the top and destructive interference on the bottom \cite{clive_pfr,clive_rsi}.
This is a natural consequence of a change of a strong change in the
$E\times B$ shearing rate by the bursting mode. 

Integrating over all velocity components gives a spatial profile of
fluctuation amplitude, and, the bottom and top profiles are averaged
and plotted as a function of time, together with the average phase
velocity profiles in Fig. (\ref{fig:pciprofs_1_2}c,d). It is clear
that there are several rapid changes in the phase velocity, large
enough to shift the rotation from the ion to the electron diamagnetic
direction. The turbulence amplitude is bursting up and down in this
interval, and there is sufficient turbulence signal at all times to
characterise the phase velocity. In fact, during the bursts, a slight
reduction in the turbulence amplitude occurs. This effect will be
analysed in more detail later. In Fig. (\ref{fig:pciprofs_1_2}e),
the Logarithm of ratio of the fluctuation amplitude on top to that
on the bottom is plotted, conveying how the eddy tilt angle changes
dramatically during these bursts as discussed above. 

These dramatic changes in the rotation velocity are likely to be due
to a localized change in the radial electric field, and appear to
have the signature of EICs, which are associated with fast ion losses
and/or redistributions (shown in \cite{eic_nf} by signatures in the
HIBP and CXRS systems). The magnitude of the change in velocity with
each burst is demonstrated more clearly using a narrow time interval
in Fig. (\ref{fig:diftime}a) (where the phase velocity is offset
from a particular time in-between the bursts), and compared the fluctuation
amplitude profile evolution in Fig. (\ref{fig:diftime}b), as well
as the a time evolution of the magnetic probe signal and wavelet spectrogram
in Figs (\ref{fig:diftime}c,d). It is clear that the phase velocity
(electric field) changes are clearly localized in the region $\rho=0.6-0.9$,
and turbulence amplitude is reduced in that time, though the asymmetry
changes. These changes are broadly similar to the magnitude of the
electric field change observed for an EIC, and is in the same negative
radial direction (poloidally electron diamagnetic direction). In \cite{eic_nf},
it was shown that for EICs, the radial current associated with the
outward movement of fast ions could be derived from the time derivative
of the electric field according to Faradays equation $\nabla\times B=j+\epsilon dE/dt$
(assuming B doesn't change). It was argued that one can determined
the loss current from the rate of change electric field in the build
up phase of the event. On the other hand, in the recovery phase, the
phase velocity decreases on a timescale related to the parallel-viscosity
(shown in Fig. (\ref{fig:diftime}a)) \cite{eic_nf}.

\subsection{Magnetic probe signatures and mode identification\label{subsec:Magnetic-probe-signatures}}

\begin{figure}
\includegraphics[width=8cm]{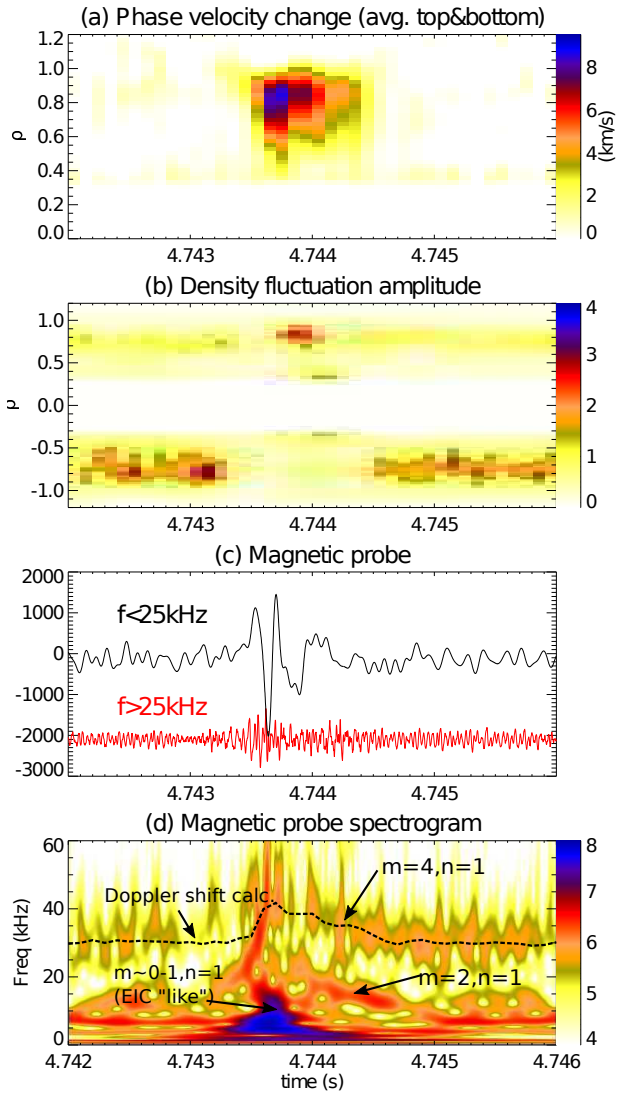}

\caption{(a) Radial distribution of the change in PCI phase velocity (averaged
over top and bottom) from the initial reference time for He-rich discharge
\#128670, compared with (b) the fluctuation amplitude, resolved over
the top and bottom. (c) The magnetic probe ($\dot{b}_{\theta}$) waveforms
for frequency intervals above and below 25kHz, and (d) magnetic probe
wavelet spectrogram ($6^{th}$ order Morlet). Arrows indicate the
mode numbers for the respective mode branches, and the line indicates
the Doppler shift frequency change (from the initial frequency of
30kHz) calculated using the phase velocity change in (a).\label{fig:diftime}}
\end{figure}

The magnetic probe signals (which measure $dB_{\theta}/dt=\text{\ensuremath{\omega\tilde{B_{\theta}}}}$)
show clear signatures of these bursts. In Fig (\ref{fig:diftime}c),
the raw signal is shown for the He rich discharge, for one such burst
event, separated out into high and low frequency components (above/below
25kHz), and in Fig. (\ref{fig:diftime}d), which shows a wavelet spectrogram.
Low frequency components around 8kHz are present before the event,
characteristic of resistive-interchange type activity. At the time
of the burst, its amplitude increases and frequency broadens, but
remains very transient, dying away after \textasciitilde{}200$\mu$s.
This is particularly evident in the low-pass filtered raw signal in
Fig. (\ref{fig:diftime}c), which shows an increasing amplitude, but
the peak remains there for only one cycle of the oscillation. Analysis
of the Mirnov array indicates that this mode has a clear n=1 structure,
and a very low m number, between 0 and 1, which is similar to that
observed at the onset of an EIC mode. 

Both during and between events, a persistent (though bursty) mode
around 30kHz is present. At the time of the bursts, this mode jumps
up in frequency (to 40-50kHz) and amplitude followed by a chirp-down
type behaviour. The toroidal mode number of the higher frequency 30-40kHz
mode is found to be n=1, whilst the poloidal mode number is found
to be m=4, and it propagates in the electron diamagnetic direction
poloidally and clockwise direction toroidally (same as EIC). Furthermore,
the poloidal propagation speed $2\pi fr/m$ is roughly the same as
that for the EIC since both the frequency and poloidal mode number
are 4 times that for the EIC. The chirping up then down behaviour
may be an influence of the change in plasma rotation velocity. Given
that the PCI phase velocity is close to the plasma $E\times B$ rotation
velocity, the change in mode frequency can be computed from the simple
formula $\Delta f=\Delta v_{phase}m/(2\pi r_{eff})$ (neglecting changes
in the toroidal rotation). This frequency is computed, offset from
an initial value of 30kHz, and compared with the mode frequency in
Fig. (\ref{fig:diftime}d). It is clear that this waveform, indicated
by dashed line, closely matches the changes in mode frequency. However,
there may be additional corrections owing to the intrinsic mode frequency.
It has been suggested that this could be in the electron-diamagnetic
direction \cite{nishimura_sim_eic_2016}. 

Conventionally, if the mode were resonant at a rational surface with
the resonant surface with $\iota=\iota_{*}$, that value would be
given by $\iota_{*}=n/m=0.25$. However, the minimum value of $\iota$
is $\approx0.4$, occurring in the centre of the plasma. As the $\iota=0.25$
surface is not present in the plasma, it suggests that this mode must
have considerable $k_{||}$. One possible explanation is that the
mode is like Alfven wave in 3D toroidal geometry, where gap locations
can occur for a range of couplings to the poloidal and toroidal field
harmonics $(\mu,\nu)$ (with $N=10$ being the toroidal field periodicity
of the Heliotron) \cite{ko2003}:

\begin{equation}
\iota_{*}=\frac{2n+\nu N}{2m+\mu}.
\end{equation}

Considering possible values for field harmonics $(\mu,\nu)$ as (2,1)
and (1,1) for ``helical ripple'', we can obtain $\iota_{*}=1.2,1.33$
respectively, which are located at $\rho=0.9,0.95$ respectively.Such
couplings would put this mode into the family of the ``helicity-induced
Alfven eigenmode'' (HAE). Previously reported HAE modes were of higher
frequency (in the range $\approx$200kHz), as reported in \cite{yamamoto_hae},
but gap mode solutions can exist over a broad range of frequencies.
Furthermore, the mode frequency is broadly in in the range of $\beta$-induced
Alfven eigenmode (BAE) and the energetic particle GAM \cite{egam},
but since $v_{A}\sim1/A$, the the frequency should change with He
concentration, which is not the case. To determine what fast particles
destabilise this mode, we have to consider the condition \cite{ko2002}: 

\begin{equation}
f-(m+j\mu)f_{\theta}+(n+j\nu N)f_{\phi}=0,
\end{equation}
(with $j=0,\pm-1$), and $f_{\theta}$, $f_{\phi}$ being the poloidal
and toroidal precession frequencies. It is not possible for perpendicularly-injected
beam ions to satisfy this resonance condition, despite having the
same poloidal propagation velocity as the EIC, which is driven by
these particles \cite{eic_nf}, because the mode propagates in the
electron diamagnetic direction, whilst these fast ions precess in
the ion diamagnetic direction. It was shown that for an EIC, coupling
with field harmonics $(\mu,\nu)=(0,1)$ could create such a resonance
in the ion diamagnetic direction, however no such condition can be
found for the 30kHz mode. Such a discrepancy has also been reported
before. Reverse shear Alfven eigenmodes modes in LHD also propagated
in the electron diamagnetic direction contrary to expectations \cite{toi_2010}.
Whilst this mode does not appear to fit a resonance condition, the
fact that is amplitude increases during EIC bursts, and the fact that
the mode on-set occurs after perpendicular NBI injection as shown
in Fig. (\ref{fig:nbicor}b), suggests that it is driven by perpendicularly-injected
fast ions. Whilst the nature and drive of the mode remains elusive,
and further investigation is required to fully understand it, the
EIC mode amplitude is distinctly greater than that of the 30kHz mode.
From the mode amplitudes presented in Fig. (\ref{fig:diftime}), the
magnetic probe (which measures $dB_{\theta}/dt$) for the low frequency
EIC mode is $\sim2\times$ that of the 30kHz mode. Considering the
frequency dependence of the probe response, this means that the field
perturbation would be around $10\times$ larger. As shown later in
Fig. (\ref{fig:br_vr_mag}c,d), in other discharges, the EIC mode
amplitude is even higher relative to the 30kHz mode.

\subsection{Generation of bursts by Perpendicular NBI }

\begin{figure}
\includegraphics{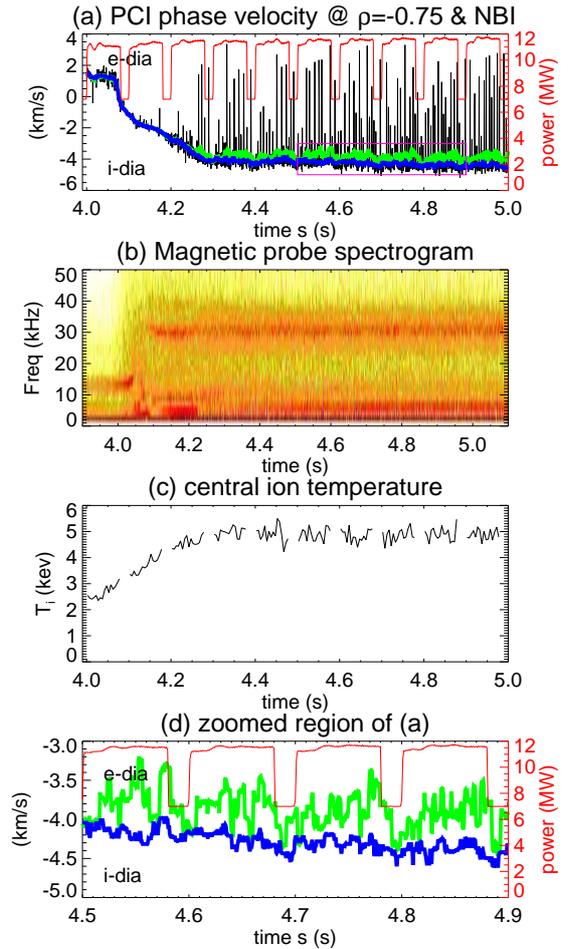}

\caption{For 27\% He concentration shot \#128665: (a) Time history of phase
velocity (black), as well (blue) as a running median filter over 10ms
(blue), to remove spike features, and (green) a boxcar average filter,
compared with perpendicular beam power (red). (b) Magnetic probe spectrogram.
(c) Central ion temperature. (d) zoomed region of (a) to show how
average phase velocity (green) reduces when beam power reduces\label{fig:nbicor}}
\end{figure}

The conditions which lead to the EIC bursts and the 30kHz mode can
be examined by looking over a wide time interval. The phase velocity,
together with the perpendicular NBI \#4 current waveform, as well
as the magnetic probe spectrogram and central ion temperature for
the 27\% He shot \#128665 in Fig. (\ref{fig:nbicor}). At t=4.0s,
the perpendicularly-injected beam power increases from 7 to 12MW and
is then modulated at 10Hz with a 80\% duty cycle. The ion temperature
starts to increase and reaches its equilibrium level at $t=4.3$s.
During the time interval $t=4-4.3$s, the phase velocity at $\rho=0.75$
switches from the electron to ion diamagnetic direction. The bursting
activity starts to commence around 4.1s although they start out small
and do not reach their maximum magnitude until around 4.3s.

On the other hand, the 30kHz mode appears at t=4.1s. Given that the
fast ion slowing down time is around 100ms, the 30kHz mode is triggered
once the fast ions start to come into equilibrium with the bulk ions,
but before the ion temperature has reached its maximum level. Thus
we can conclude that the 30kHz mode is either destabilized by the
ion temperature gradient, or by fast ions of energies and pitch angles
which are different from the the birth injection and pitch angle,
thus taking a slowing/scattering timescale in order to trigger the
mode.

It also appears from a glance from Fig. (\ref{fig:nbicor}a) that
the phase velocity bursts are smaller and less regular in the 20ms
intervals when the beam power is lower. Also indicated in Fig. (\ref{fig:nbicor}a),
a running median filter over 10ms is indicated in blue, which removes
the bursting velocity features, to give a measure of the baseline
variation of the phase velocity, as well as aboxcar average filter
in green, which smooths over the spike features, but does not remove
them. These traces are reproduced over a narrower region for clarity
in Fig. (\ref{fig:nbicor}d) to see their correlation with the NBI
power. The difference between the green and blue curves demonstrates
the time evolution of the bursting events. From these traces, it is
clearer to see that bursting events reduce in the beam-off periods,
and how the burst rate steadily increases in time when the beam is
on.

\begin{figure}
\includegraphics{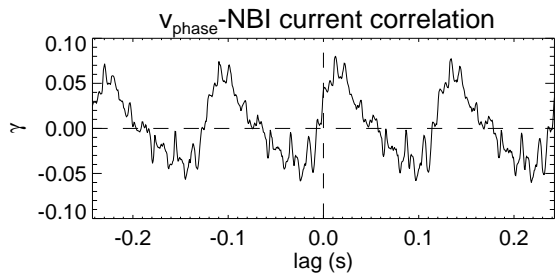}

\caption{Cross-correlation function between the perpendicular NBI current,
and the phase velocity, (minus a running median filter) demonstrating
how the bursts in the phase velocity are also modulated by the NBI
current.\label{fig:Cross-correlation-function-betwe}}
\end{figure}

The cross-correlation function between the phase velocity (corrected
with the blue baseline in (a)) and perpendicular NBI current signals
is plotted in Fig. (\ref{fig:Cross-correlation-function-betwe}).
The waveform of the correlation function has a period precisely 0.1s
which is the NBI modulation period, demonstrating again how the bursts
evolve in time. Almost identical correlation signatures were found
in other He concentration shots. This indicates that the bursts, which
is suggesting fast ion loss, is strongly affected by the perpendicular
beam power. 

\subsection{Fast ion loss measurements\label{subsec:Fast-ion-loss}}

An examination of the response of various fast ion diagnostic systems
(summarized in \cite{fastparticle_diag_review}) may confirm whether
this bursting mode is associated with fast ion redistribution/losses.
Several neutral particle analysers are available, which collect CX
neutrals from the ambient neutral density. A Compact Neutral particle
analyser measures the energy spectrum of such neutrals, being sensitive
mostly to hydrogen \cite{cnpa}. The sightline for this diagnostic
is vertical, mostly perpendicular to the field. As such, this is mostly
sensitive to fast ions arising from perpendicular beam injection (NBI
4\&5). Also, as shown later by spectra in Fig (\ref{fig:cnpa_cmp}),
signal is present at energies above the injection energy of the perpendicular
beams (40keV), which is considered to be due to slowing down and pitch
angle scattering of fast ions driven from the tangential injected
NBIs (1,2,3) which have an injection energy of 180keV. A silicon FNA
type detector (SiFNA) is also available which does not provide any
energy resolution \cite{sifna_osakabe}, and for these discharges
provided spectra also from only a perpendicular sight line (however,
in general, parallel sight lines are available). The radial location
from which the signal arises with both of these passive NPA diagnostics
is determined by the competing effects on one hand of the decreasing
fast ion population towards the edge of plasma, and the decreasing
neutral donor population towards the core of the plasma. Additionally,
an RF spectrometer mounted on a manipulator measures ion cyclotron
emission (ICE) \cite{ice_reference} and its very high order harmonics
\cite{ice_kstar}, and is sensitive to the perpendicular energy of
fast ions lost near the edge of the plasma.

\begin{figure}
\includegraphics{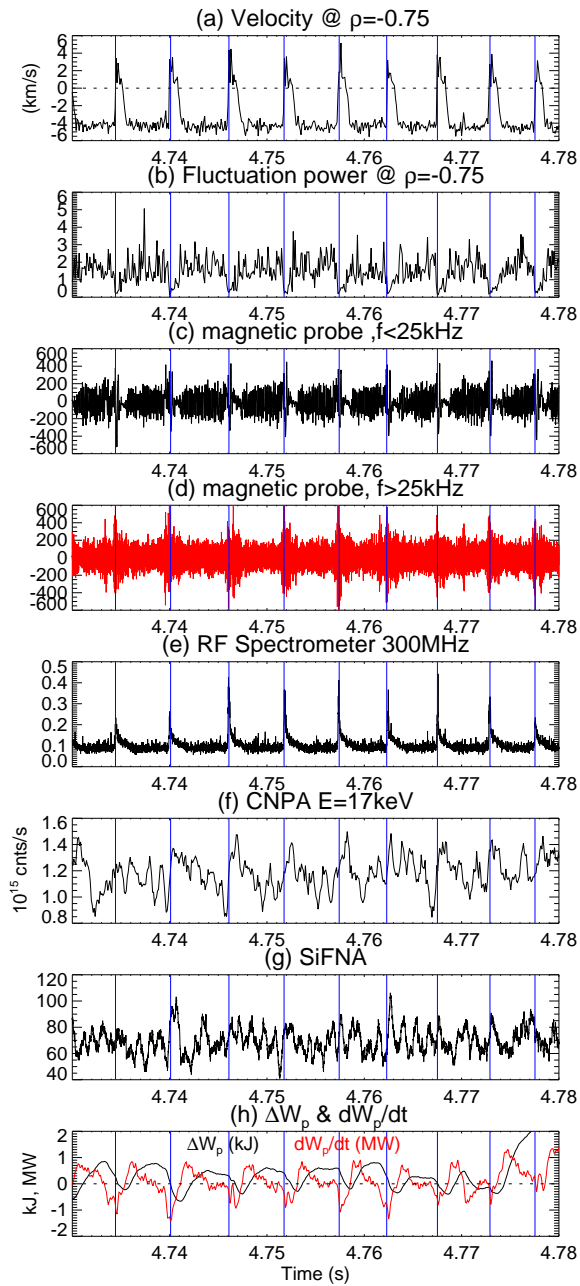}

\caption{Time history of (a) fluctuation phase velocity and (b) amplitude at
$\rho=-0.75$, compared with the raw magnetic probe signal ($dB_{\theta}/dt$),
low pass filtered (c) and high pass filtered (d) above 25kHz, (e)
the RF ICE spectrometer with 300MHz energy channel, (f) the CNPA signal
at 17keV, (g) the SiFNA signal and (h) changes of the diamagnetic
stored energy (high pass filtered above 50Hz to remove the baseline
variation), together with the time derivative (red) for the $30\%$
He shot \#128708.\label{fig:th_pci_fi}}
\end{figure}

A time history of the amplitude of these 3 fast ion diagnostics, as
well as diamagnetic stored energy and magnetic probe signal separated
out into components above and below 25kHz are compared with the fluctuation
phase velocity and amplitude at $\rho=0.75$, for the 30\% He shot
in Fig. (\ref{fig:th_pci_fi}). This simple line profile comparison
creates a clear impression of the nature of these bursts. With each
burst, the phase velocity changes suddenly in less than 0.1ms and
reverses for about 2ms. During that time, the fluctuation amplitude
decreases dramatically. At that time, the RF spectrometer channel
at 300 and 400MHz has a clear increase in the signal. This diagnostic
is considered to measure fast ions in the very periphery of the plasma,
and thus conveys the redistributed (radially transport) fast ions
to the edge of the confinement region of ``lost'' fast ions. Due
to the complex nature of ICE emission, it is difficult to identify
the position of this emission based on cyclotron frequency alone.
Nor is not clear from that exactly how many fast ions are being lost.
The CNPA and SFNA signal also have apparently positive spikes at the
time of these bursts, however their correlation is much less obvious
than with the ICE spectrometer. Nonetheless, a positive spike may
be due to the redistribution of fast ions further towards the edge
of the plasma where the donor neutral density is larger, or it may
be as a result of increased background neutral density, as a consequence
of the burst itself – which is possible considering there is a slight
drop of edge electron temperature, as shown in Sec. (\ref{subsec:Turbulence-and-heat}),
and no clear change in the electron density from the FIR interferometer. 

\begin{figure}
\includegraphics[width=8cm]{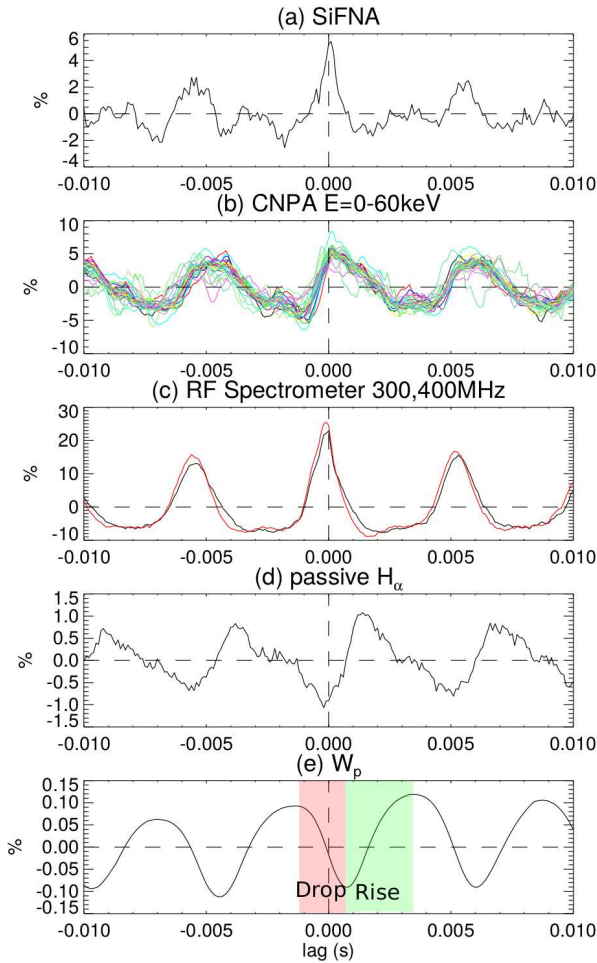}

\caption{Cross-correlation measure using $\gamma^{rel}\times100\%$ between
PCI phase velocity and (a) SiFNA, (b) CNPA energy channels (in different
colours) up to 60keV (larger energies are dominated by noise), (c)
the ICE RF spectrometer in the 300,400MHz channels, (d) passive $H_{\alpha}$,
and (e) the diamagnetic stored energy $W_{p}$. Discharge $\#127708$,
$30\%$ He.\label{fig:cc_anderes}}
\end{figure}

To investigate this phenomenon in more detail, the cross-correlation
$\gamma^{rel}$ (as defined in Eq. (\ref{eq:gammarel})) between the
PCI phase velocity and the fast in diagnostics including the SiFNA,
the CNPA and the RF spectrometer were evaluated in Fig. (\ref{fig:cc_anderes}a-c).
This analysis effectively averages over multiple events to detect
the relative influence that these events have on the fast ion signals.
Different energy channels for the CNPA are plotted, and, although
the signal changes over orders of magnitude as a function of energy,
as discussed later in Sec. (\ref{subsec:Compact-Neutral-particle}),
the relative change is independent of energy, increasing by about
10\% with each burst, up to an energy of about 60keV at which point
the SNR is not adequate. The initial conclusion from this graph is
that fast ions are lost over a broad energy range, which is different
to that shown in \cite{eic_nf} where there was a distinct change
in the character of the bursts around $E_{b}=35$keV corresponding
to the injection energy of the perpendicular beams. However, it could
also be an effect of changing neutral density. The cross-correlation
with passive $H_{\alpha}$ monitors, shown in Fig. (\ref{fig:cc_anderes}d)
indicates that the passive neutral density goes up after the burst
by about 1\%, which is insufficient to explain the increase in the
NPA signal. Therefore, radial fast ion redistribution must occur over
a broad range to regions with higher donor neutral density.

Whilst none of the above signals convey the absolute magnitude of
the amount of fast ions lost for each event, the diamagnetic loop,
conveying the stored energy in the field perpendicular direction,
is well calibrated, and is sensitive to the beam ions of NBI 4 and
5 which dominantly produce energetic particles with high perpendicular
velocity. Moreover, it does have fast enough time resolution to see
the influence of every burst event. The time history of the stored
energy $W_{p}$, high pass filtering features slower than 20ms, together
with $dW_{p}/dt$ is plotted in Fig. (\ref{fig:th_pci_fi}h). During
the burst, $dW_{p}/dt$ drops to values around -1MW, which is a significant
fraction ($25\%)$ of the $\approx0.3\times13$MW input from the perpendicularly-injected
NBI systems. However, as the burst duration of $\sim2$ms is much
shorter than the $\sim100$ms slowing down time, the overall effect
on fast ion losses is reduced. The drops are of order of 1kJ, which
is $\approx$0.1\% of the \textasciitilde{}1MJ of total stored energy.
From simulations, the total fast particle energy in these shots is
comparable to the thermal energy. Therefore, each burst on its own
is unlikely to cause enough energy loss to explain the lower $T_{i}$
in the H-rich shots. However, if the burst events are not immediately
redistributing the ions outside the plasma volume, the change in $W_{p}$
may only be slight as it is a globally-averaged quantity. In order
to understand clearly the evolution of the stored energy in the burst
and recovery phases, another cross-correlation calculation is performed
between the PCI phase velocity and the stored energy, plotted in Fig.
(\ref{fig:cc_anderes}e). This figure tends to give the average evolution
of the stored energy in the recovery phase. This clarifies that there
are only two phases: the drop in energy associated with the burst,
then the rise in stored energy after the MHD event has finished. However,
there is no significant period where the stored energy remains constant,
indicating that the stored energy never completely ``recovers''
after each burst. Therefore, although each bust only loses $\sim$1kJ,
the combined effect of all these bursts may reduce the stored energy
by more than that value. Analysis of $W_{p}$ is not completely conclusive,
as changes during each burst may also be associated with changes in
the bulk ion and electron density and temperature profiles. 

\subsection{Ion species dependence of bursts and estimation of loss current}

\begin{figure*}
\includegraphics{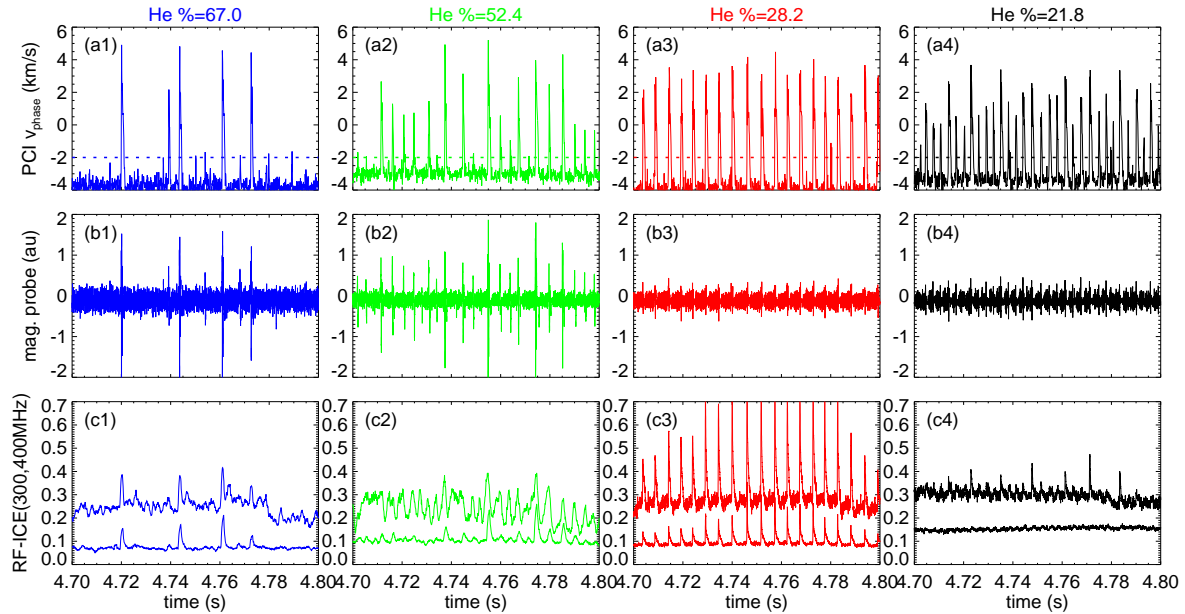}

\caption{Comparison of the temporal evolution of (a) the phase velocity at
$\rho=0.75$, as well as (b) the low frequency component of filtered
magnetic probe signals (\textless{}25kHz), and (c)d RF ion cyclotron
emission spectrometer channels at 300,400MHz, showing how the the
bursting behaviour varies with He fraction (1-4). The critical velocity
$v_{crit}$ defining a burst is indicated with the dashed line in
(a).\label{fig:burstcmp_ispec}}
\end{figure*}
The bursting-MHD features described are more frequent in more H rich
discharges. Phase velocity time traces at $\rho=0.75$ are compared
between each of the four shots discussed in Sec. (\ref{sec:Ion-ITB-comparison-discharges})
in Fig. (\ref{fig:burstcmp_ispec}a1-4), together with low-pass filtered
(f\textless{}25kHz) traces of the magnetic probe signals, to capture
the EIC events in Fig. (\ref{fig:burstcmp_ispec}b1-4), and traces
of the RF ICE spectrometer at 300 and 400Mhz in Fig. (\ref{fig:burstcmp_ispec}c1-4).
It is clear that in the He-rich discharge, the bursts only occur ever
20-30ms, whilst in the H-rich discharge, the bursts occur roughly
every 4ms. This suggests that the fast particle population and fast
particle beta, which drive the modes is substantially different in
each of the four cases. Moreover, these changes are unlikely to be
due to any changes in the bulk plasma properties, as the $n_{e}$
\& $T_{e}$ profiles are near identical in all cases, and the $T_{i}$
profile changes as shown in \cite{nagaoka}. It was shown in that
paper that the ion heating power per particle was also unchanged between
these shots. 

For all discharges, there are signatures of the losses in the RF ICE
spectrometer in Fig. (\ref{fig:burstcmp_ispec}c1-4), however these
vary considerably in shape and magnitude. This may reflect a change
phase-space redistribution of fast ions relative to the sensitivity
of the diagnostic. Also, the RF emission intensity may not necessarily
be linearly proportional to the energetic particle loses. Bursts in
the magnetic signals are always perfectly correlated those in the
phase velocity, even amongst these four discharges. Furthermore, in
He-rich discharges, the bursts in the magnetic probe signals are considerably
stronger, whilst the change to the phase velocity is around the same
value. The longer period between events in He suggests that the modes
are more ``stable'', however the magnetic probe signatures are larger,
with the same resulting fast ion losses, assuming that they can be
conveyed by the PCI velocity. 

In all cases, the duration of the bursts is roughly constant at around
2ms. However, the magnitude of the bursts in phase velocity (i.e.
the fast-ion driven radial electric field) are not always the same,
more particularly for the 21.8\% and 52.4\% He shots. To quantify
the details of the PCI phase velocity change in more detail, the relative
burst rate is computed, defined by:

\begin{equation}
r=1/T\int_{\Delta v_{phase}>v_{crit}}dt,\label{eq:br}
\end{equation}

where $v_{crit}$ is arbitrarily defined (here at -2km/s) to distinguish
these MHD events, and T is the time window length (here we use 0.1s).
If we assume that bursts represent fast ion losses, then a quantity
related to the average fast ion loss current can be evaluated the
integral of the change in phase velocity:

\begin{equation}
j_{loss}=1/T\int_{\Delta v_{phase}>v_{crit}}\Delta v_{phase}dt,\label{eq:jr}
\end{equation}

assuming that the losses are all dependent on the magnitude of the
initial change in $v_{phase}$ (during the first 0.1ms of the burst),
and that the $\sim2$ms timescale for the relaxation is the same in
all cases. A precise evaluation of the loss current, considering the
physics of momentum transport and fast ion redistribution is beyond
the scope here, but has been touched upon in \cite{eic_nf}.

The average change of phase velocity for each burst $\Delta v_{avg}$
can be calculated from 

\begin{equation}
\Delta v_{avg}=j_{loss}/r
\end{equation}

\begin{figure}
\includegraphics{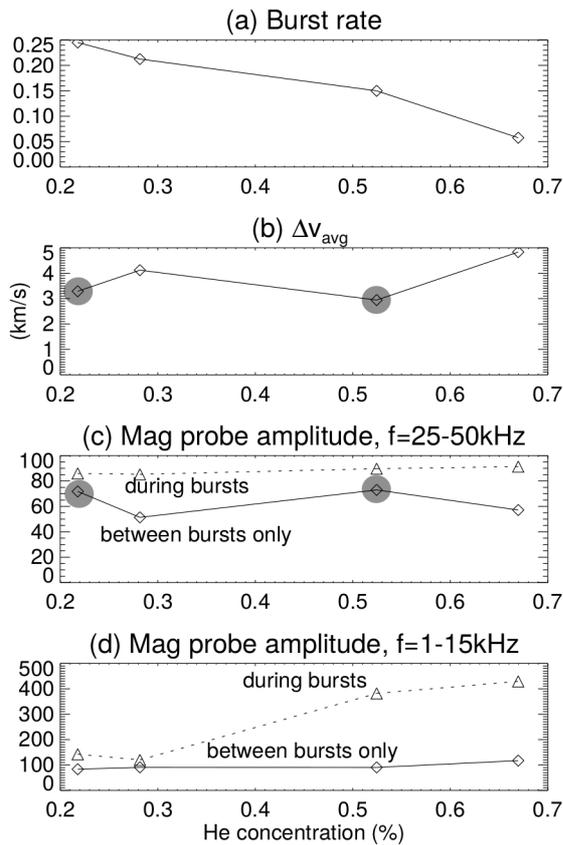}

\caption{(a) The burst rate (defined as the fraction of time for which $v>v_{crit}$,
evaluated over 4.5-5s, calculated from the time history of the phase
velocity at $\rho=-0.75$ as shown in Fig. (\ref{fig:burstcmp_ispec}))
plotted as a function of He concentration. (b) the average velocity
of each burst, $\Delta v_{avg}$, calculated as a function of the
He concentration. Average amplitude of the magnetic probe signals
for 25-50kHz (c) and 1-15kHz (d), as a function of the He concentration,
analyzed during time windows during bursts, and in between bursts.
The points in grey highlight the correlation between $\Delta v_{avg}$
and high frequency mode amplitude between bursts. \label{fig:br_vr_mag}}
\end{figure}

The burst rate and $\Delta v_{avg}$ are plotted vs. He concentration
in Fig (\ref{fig:br_vr_mag}a,b), whilst the dependence of $j_{loss}$
is discussed in Sec. (\ref{subsec:Role-of-EIC}). It is clear that
the burst rate is inversely related to the Helium concentration, however,
the values of $\Delta v_{avg}$ are lower for the 21\% and 53\% He
concentration discharges. One possibility for this is the additional
viscosity/damping to the rotation in the presence of magnetic perturbation,
which can introduce higher harmonic field ripple \cite{ida_viscosity_1997}.
To investigate the role of this perturbation, the amplitude of the
magnetic probe signals, filtered between 25-50kHz (to obtain the component
with 30kHz and associated chirping features associated with m=4,n=1),
averaged over times both with bursts and without bursts are shown
in Fig (\ref{fig:br_vr_mag}c). It is clear the magnetic probe amplitudes
in between bursts are higher for the $21\%$ and $53\%$ Helium discharges,
which are precisely the ones with smaller $\Delta v_{avg}$. Also
in this figure, it is observed that the amplitude of 25-50kHz fluctuations
is higher during the bursts but with little dependence on the He concentration. 

Finally, in Fig. (\ref{fig:br_vr_mag}d), the magnetic probe amplitude
for low frequency components (f\textless{}25kHz) is indicated, which
tend to encapsulate the EIC-like burst features ($m\sim0-1$,$n=1$),
both during and between bursts. It is clear that the EIC mode amplitude
increases markedly with He concentration, being near the inter-burst
level for H rich discharges, and increasing to a level 5 times the
value in between bursts in He rich discharges. This behaviour is somewhat
``anti-correlated'' with the burst rate dependence indicated in
Fig. (\ref{fig:br_vr_mag}a). This is invaluable information to understand
the drive of the EIC: in He discharges, the mode is more ``stable'',
in that the modes are driven less frequently, however when an event
occurs, it is larger. However, the fast ion losses for each event,
characterised by $\Delta v_{avg}$, are largely unchanged (excepting
for the small reduction in the 21\% and 53\% discharges as discussed).

\subsection{Turbulence and heat transport modulation due to bursts\label{subsec:Turbulence-and-heat}}

During the burst events, as has been shown, the velocity reverses,
producing strong $E\times B$ velocity shear. Associated with this,
the density fluctuations decrease dramatically for that period of
time. In \cite{eic_nf}, it was shown that strong EIC events were
associated with a transient improvement of particle and both electron
and ion thermal confinement, and this is certainly the case for electron
thermal confinement in these discharges, as shown below. On the other
hand, the average level of turbulence hardly changes between H and
He discharges \cite{tanaka_nf2017}, despite the improvements during
the bursts in H. This complex interplay therefore requires more detailed
analysis. Whilst it would be informative to understand the correlation
with ion temperature, the SNR of the CXRS data is not sufficient to
make any conclusion. On the other hand, the ECE system shows clear
changes in the electron temperature which can be analyzed.

\begin{figure}
\includegraphics{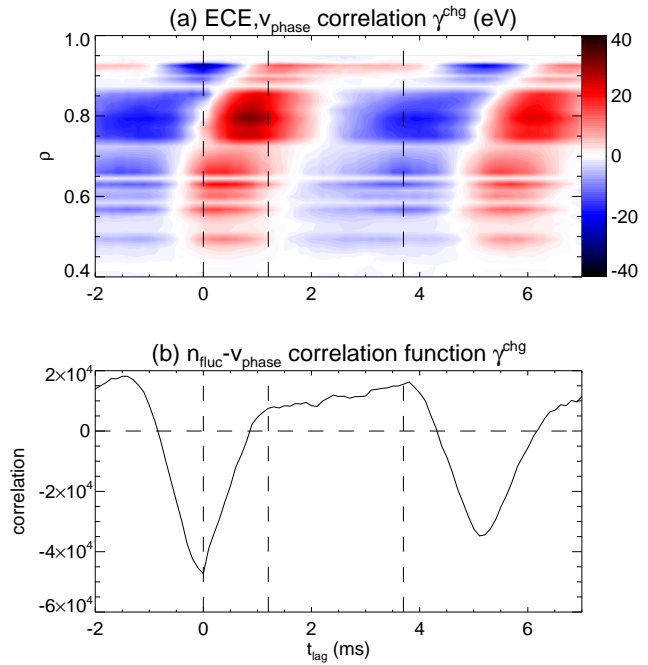}

\caption{Normalized correlation $\gamma^{chg}$ (from eq. (\ref{eq:gammachg}))
between $v_{phase}$ and (a) ECE signals as a function or $\rho$
and delay time $t_{lag}$, and (b) $\tilde{n}$, for 30\% H shot \#128708.
\label{fig:ececorr}}
\end{figure}

The electron temperature, measured using electron cyclotron emission
(ECE) diagnostic, shows distinctive changes during the burst events.
To clarify this, the normalised correlation function $\gamma^{chg}$
is calculated in Eq. (\ref{eq:gammachg}), indicating on average,
the change in value of temperature with each burst. This is indicated
in as a 2D plot as a function of delay time and normalised radius
in Fig. (\ref{fig:ececorr}a) for the 30\% H rich discharge. There
are clear temperature changes associated with the bursts, including
a temperature drop around -20eV associated with the burst (at $\rho=0.9,$
$t_{lag}=0$), and a maximum temperature rise around$\sim$30eV at
$\rho=0.8,$ $t_{lag}=1$ms. The increase in the temperature is suggestive
of a transport improvement, possibly generated by the increased velocity
shear. According to Fig (\ref{fig:diftime}a) the change in $E\times B$
rotation is largest around change around $\rho=0.8$. so that the
region $\rho=0.6-0.8$ is where the strongest velocity shear is situated.
This corresponds well with the region of largest temperature increase
after the event. After the bursting event has ceased, the temperature
decreases, and this excess heat tends to propagate outwards to $\rho=0.9$
at $t_{lag}=1-3$ms as shown in Fig. (\ref{fig:ececorr}a). Thus,
the apparent temperature ``drop'' at $t_{lag}=0$ is more a consequence
of the delayed propagation of the heat pulse from further in.

The cross-correlation $\gamma^{chg}$ between the phase velocity and
the fluctuation level (both at $\rho=0.75$) is plotted for the 30\%
discharge in Fig. (\ref{fig:ececorr}b). It is clear that the drop
in fluctuation amplitude, between $t_{lag}=-1$ to 1ms is well correlated
with the $T_{e}$ rise at $\rho=0.8$, and the ``flat top'' increase
in fluctuation-amplitude is correlated with decreasing electron temperature
at $\rho=0.8$. While the average central electron temperature does
not change with He concentration, there is in fact some transient
change with each burst, but it is more concentrated towards the edge.
It is worthwhile to keep in mind that for these discharges, it was
shown that the turbulence the amplitude is well correlated with ion
temperature gradient \cite{tanaka_nf2017}, consistent with it being
in the ITG regime. We also note that it was shown that for EICs \cite{eic_nf},
both ion and electron thermal conductivity was improved after each
burst, purely showing that the two are connected, presumably as both
are affected by $E\times B$ shear.

The average turbulence level is the same in the H and He-rich discharges
\cite{tanaka_nf2017}, and, because of the larger occurrence of bursts
in H-rich discharges, where the fluctuations are suppressed, the fluctuation-level
in-between bursts is in-fact higher. Comparing the electron temperature
in Fig. (\ref{fig:ececorr}a), and turbulence amplitude in figure
(\ref{fig:ececorr}b), the turbulence amplitude increases up to a
delay of 4ms after the burst event, when the ``heat pulse'' has
propagated towards the edge, thereby increasing the electron temperature
gradient (at some location) and thus the ``potential'' drive for
the turbulence (strictly, the ion temperature gradient is expected
to have the most dominant influence). However, it is expected that
a similar improvement may occur to the ion temperature, as this was
reported for an EIC event in \cite{eic_nf}. The details of these
bursts allows to understand the critical gradient behaviour of turbulence
as predicted by gyro-kinetics \cite{nunami_cmp_h_he}.

\section{variation of confined fast ion density}

The time-averaged value of various measures of the fast-ion distribution
function may give further clues as to the reason for the ion temperature
dependence on He concentration. This may resolve the issue of understanding
the magnitude of the fast ion losses from the MHD events, or may indicate
how other processes, such as neutral penetration into the plasma volume,
and charge-exchange losses, play a role.

\subsection{Compact Neutral particle analyser\label{subsec:Compact-Neutral-particle}}

The NPA systems which were utilised in Sec. (\ref{subsec:Fast-ion-loss})
including the CNPA and the SiFNA can convey information also about
the average level of fast ions in the He/H comparison discharges.
The complicating factor, however, is that the effect of the difference
of the cross-sections between fast hydrogen ions and background hydrogen
or background Helium atoms has to be considered, as well as the variation
in the background neutral density. Whilst this may seem complicated,
a simple analysis of the energy-resolved CNPA system can give a clue
as to the variation on the fast ion density in between the H and He-rich
discharges. 

\begin{figure}
\includegraphics{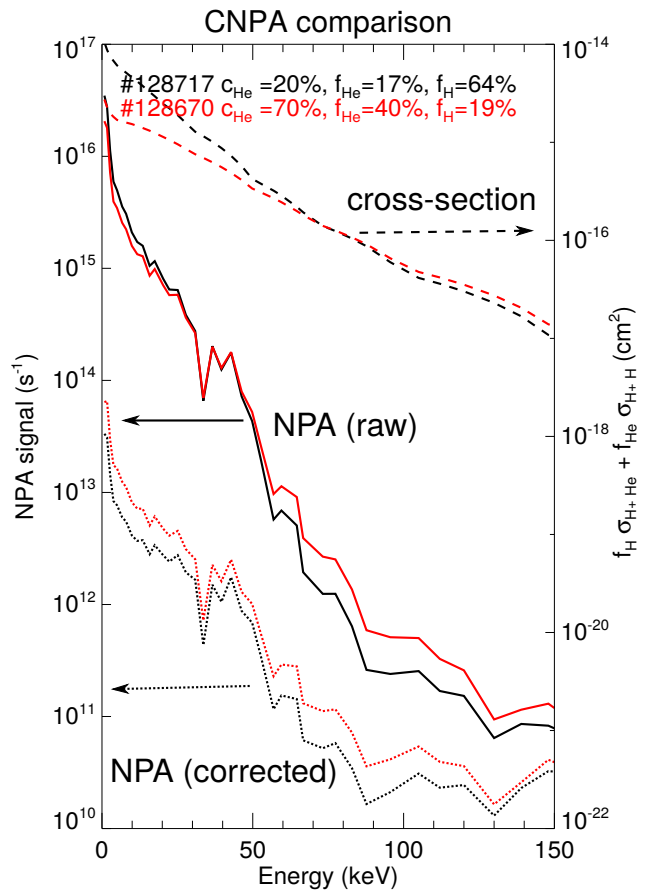}

\caption{Solid line: CNPA Energy spectra for the interval 4.7-4.9s for the
He and H rich discharges (red,black). Dashed line: species-weighted
charge-exchange cross-section$\sigma_{H-H^{+}}(E)f_{H}+\sigma_{He-H^{+}}(E)f_{He}$
(numbers on the right-hand axis). Dash-dot line: CNPA signal corrected
with this cross-section factor.\label{fig:cnpa_cmp}}
\end{figure}

The CNPA energy spectrum for the H and He rich shots is plotted in
Fig. (\ref{fig:cnpa_cmp}). For energies lower than 50keV, the He-rich
discharges have lower signal, whilst for E\textgreater{}50keV, H-rich
discharge has more NPA signal. The charge-exchange cross-sections
for H and He collisions with H ions has been obtained from ADAS \cite{adas}.
The H-He cross-section is lower for E\textless{}40keV, and higher
for E\textgreater{}40keV. The fast ion density $n_{fast,H}(E)$ as
a function of energy E can be related to the NPA signal $S_{NPA}(E)$
from the equation:

\begin{equation}
n_{fast,H}(E)\propto\frac{S_{NPA}(E)}{\sigma_{H-H^{+}}(E)n_{H^{0}}+\sigma_{He-H^{+}}(E)n_{He^{0}}}
\end{equation}

where $\sigma_{H-H^{+}},$ $\sigma_{He-H^{+}}$are the energy dependent
charge-exchange cross-sections, $n_{H^{0}}$,$n_{He^{0}}$ are the
background neutral density of Hydrogen and Helium (at the position
of charge-exchange). Whilst strictly, the radial dependence of all
these quantities is necessary for a thorough analysis, the known cross-section-dependent
effects can be divided out in proportion to the known fraction of
Helium and Hydrogen ion densities $f_{He},f_{H}$, as defined in Eq.
(\ref{eq:fhe},\ref{eq:fh}), which should be proportional to the
neutral densities at the periphery of the plasma. (In fact, the neutral
spectral lines are used to monitor the plasma ion density). This factor,
and the corrected CNPA spectra are indicated in Fig. (\ref{fig:cnpa_cmp}).
After normalisation, He-rich discharges have more signal than H-rich
discharges. On the other hand, as discussed earlier in Sec. (\ref{subsec:H/He-Neutral-Penetration}),
the donor neutral density profile decreases much more rapidly into
the plasma for He discharges, and so this result suggests that there
is in-fact a higher fast ion density at some radius near the edge
of the plasma (where most of the NPA signal comes from) in He discharges.

\subsection{Diamagnetic stored energy \label{subsec:Diamagnetic-stored-energy}}

\begin{figure}
\includegraphics[width=8cm]{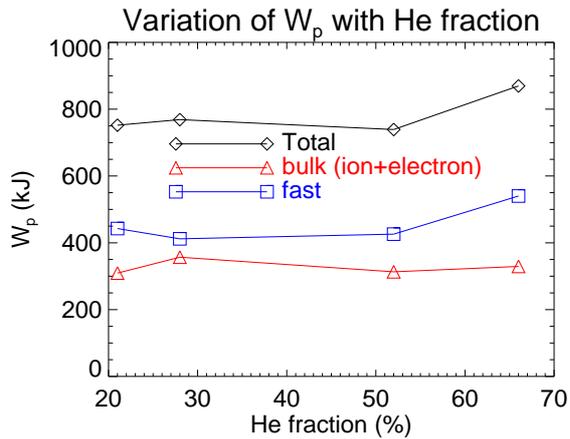}

\caption{Variation of diamagnetic $W_{p}$, thermal $W_{p}$ and inferred fast
ion component with He fraction discharge for the interval 4.5-4.9s.
Uncertainties of around $5\%$ in the thermal and therefore fast ion
stored energy due to uncertainties in the thomson and charge exchange
data, and uncertainty of the flux surface mapping. \label{fig:Variation-of-diamagnetic}}
\end{figure}

The diamagnetic loop signal varies in between the H and He rich discharges.
To infer the fast ion component, the stored energy in the bulk plasma
must be evaluated and subtracted out. For this, the Thomson scattering
system is used to determine electron density and temperature, whilst
the CXRS system is used for the ion temperature, and the ion density
is accounted for according to the uniform ratio evaluated from passive
spectroscopy. These results are plotted in Fig. (\ref{fig:Variation-of-diamagnetic}).
As the diamagnetic loop is sensitive only to the perpendicular energy,
the fast ion energy represents the perpendicularly-injected beams.
It is clear that the fast-ion component of the signal increases with
He concentration just like the NPA signal, except for the discharge
at 30\% He concentration.

\section{Physics Discussion}

\subsection{Effect of ion species on background neutral density\label{subsec:H/He-Neutral-Penetration}}

A change of working gas from H to He can produce can significantly
impact the background neutral density in the plasma region. This can
be understood by considering a simple model for neutral penetration
in plasma, which is developed for H plasmas \cite{lenhert}. In this
model, a two fluid type 'quasi-kinetic' model is employed to characterize
the penetration of neutrals into the plasma from the edge, separated
into ``slow'' neutrals, having a velocity characteristic of the
background edge neutral temperature , and ``fast'' neutrals, which
have undergone a charge-exchange reaction. The e-folding lengths for
penetration are given by slow decay length $L_{s}=v_{tn}/n_{i}\sqrt{(\xi_{iz}+\xi_{ins})(2\xi_{iz}+\xi_{ins})}$,
and fast decay length $L_{f}=v_{ti}/n_{i}\sqrt{\xi_{iz}(2\xi_{iz}+\xi_{inf})}$,
where $v_{tn}$, $v_{ti}$ is the neutral and ion thermal velocities,
and $\xi_{iz}$ is the electron impact rate coefficient, and $\xi_{ins},\xi_{inf}$
is the rate coefficient for charge exchange between ion and neutral
species for the slow and fast populations respectively. Whilst this
model is simple and neglects complex kinetic and geometrical issues
associated with the boundary plasma, it serves to illustrate the physics.
For the typical parameters of the edge plasma, taken for example with
$T_{e}=T_{i}=100$eV, $T_{n}=1eV$, and $n_{e}=10^{18}m^{-3}$, rising
to $10^{19}m^{-3}$ within 12cm \cite{nagaoka}, and obtaining the
relevant cross-sections for ionization and charge exchange from \cite{nist},
one deduces $L_{s}=$27cm (for $n_{e}=10^{18}m^{-3}$) and L$_{s}=2.7cm$
(for $10^{19}m^{-3}$), so it might penetrate realistically to a depth
$\approx\sqrt{2.7\times12}cm=6cm$. For fast neutrals, a higher density
of $10^{19}m^{-3}$ is appropriate and so $L_{f}=30$cm. The relative
abundance of fast neutrals depends on the ion density at the point
of charge exchange, which can occur most appreciably only in the last
6cm of plasma, where the density and temperature can climb to appreciable
values.

For Helium plasma, the difference is that the charge-exchange reaction
which can result in fast neutrals can only occur for singly-charged
Helium ions: $\text{\ensuremath{\mathrm{He}} + }\textrm{He}^{+}\rightarrow\textrm{He}^{+}+\textrm{He}$.
For the reaction $\text{\ensuremath{\mathrm{He}} + }\textrm{He}^{2+}\rightarrow\textrm{He}^{+}+\textrm{He}^{+}$,
there are no neutral reaction products. Considering that the ionization
potential for $\textrm{He}^{\textrm{2+}}$ is 54eV, the abundance
of $\textrm{He}^{+}$would decrease dramatically as the electron temperature
increases which would certainly be very far out from the core region.
Therefore, charge exchange and the production of fast neutrals could
only occur within the last few mm of plasma, where the ion density
and temperature are much lower (compared with Hydrogen). Consequently,
the neutral density in the core of Helium-rich plasma is expected
to be lower than in Hydrogen rich plasma.

Additionally, since these are matched electron density discharges,
the neutral density in He need only be half of that of H on account
of having two bound electrons. We also note in these discharges are
neither completely pure H or pure He, the penetration of H into the
mixed He/H discharges may contribute a significant effect. 

\begin{figure}
\includegraphics[width=8cm]{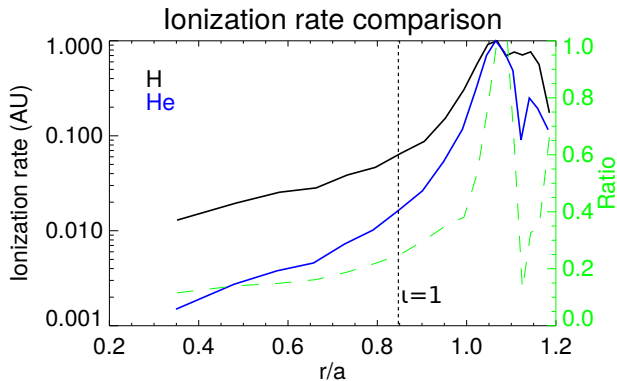}

\caption{Comparison of the ionization rate profile between H rich shot \#118352
and He rich shot \#119049 (which were ECH heated only).\label{fig:neutral}}
\end{figure}

A detailed calculation of the neutral profiles has not been undertaken
for these discharges. However, it is expected that the results published
in \cite{makinoeps} can be used to understand its effect. In that
work, the relative values of the ionization rate calculated with the
EIRENE code \cite{makino_ref6}, were compared between density and
power-``matched'' H and He rich discharges, very similar to the
discharges studied in this paper, however they were heated only by
ECH. This result is reproduced in Fig. (\ref{fig:neutral}). As the
density and temperature was almost identical, and given that the electron-impact
ionization rate coefficient for H is very similar to that for He \cite{nist},
the difference is mostly due to the neutral density. It is clear that
the He neutral density is almost an order of magnitude smaller in
He discharges compared to H discharges, and the discrepancy appears
even out towards the edge as far as $\rho=0.9$. An additional piece
of evidence that the Helium neutral density is lower in the core arises
from the thermal broadening of the neutral spectral lines. It was
shown \cite{goto} that the radial distribution of Helium could be
obtained from the broad wings of the neutral Hydrogen spectral lines,
in combination with data about the electron temperature profile. However,
observations show much narrower wings of the Helium spectral lines
indicating less charge exchange component – i.e. lower Helium density
in the core regions. 

\subsection{Influence of background-neutral charge-exchange on fast ion pressure
\label{subsec:Mode-Drive-by}}

\begin{figure}
\includegraphics[width=8cm]{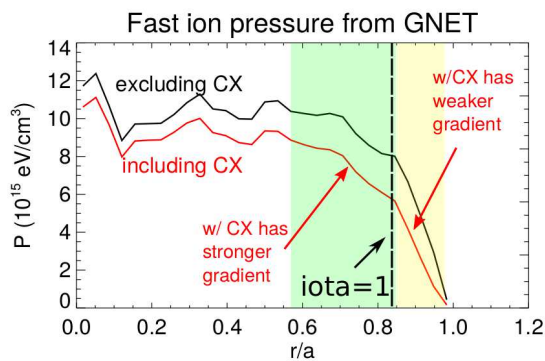}

\caption{Calculation using the GNET code of the fast ion pressure profile per
MW for perpendicular beam injection comparing the effects with and
without charge-exchange losses in the ion-ITB H rich shot \#128717.\label{fig:gnet_neutral}}
\end{figure}
The neutral density profile can have a significant impact on charge-exchange
losses of $\textrm{H}^{+}$ ions from the neutral beams. The fast
ion pressure for the perpendicularly-injected beams, calculated using
the GNET code \cite{gnet_icrh}, is plotted in Fig. (\ref{fig:gnet_neutral}),
for the H-rich discharge \#128717, both including (red) and excluding
(black) CX losses. The total stored energy from these beams decreases
by about 20\% because of charge-exchange. Surprisingly, charge-exchange
tends to decrease the fast ion pressure in the core despite the fact
that there are less neutrals there. This may be because the low-energy
neutral beams (33-43keV) tend to ionize close to the edge of the plasma,
where the background neutral density is high and thus they are susceptible
to charge exchange loss. The fast ions in the core of the plasma may
arise due to diffusion from the source region near the edge. 

As the GNET code does not presently model the background Helium neutral
distribution, nor the charge-exchange process with Helium, it is not
possible to clearly model the difference of the fast ion pressure
between H and He rich discharges. However, given that the background
neutral density is lower in Helium, the fast ion pressure will be
higher. Another contributing factor is that the charge-exchange cross-section
$\sigma_{H+He}$ is lower than $\sigma_{H+H}$ in the range 0-60keV
as shown by the dashed lines in Fig. (\ref{fig:cnpa_cmp}), contributing
to lower CX losses. This will impact on the power delivered to ions
from the perpendicularly-injected beams and to our knowledge has not
been included in the power balance analysis presented in \cite{tanaka_nf2017,nagaoka}.

\subsection{Drive and damping of fast particle MHD in He and H discharges}

The fast-ion pressure gradient should be the main drive term to determine
the instability threshold of fast ion MHD events. To explain the higher
rate of EIC events in H-rich discharges, the fast-ion pressure gradient
should be greater at the position $\rho=0.83$, where the mode is
resonant with the $\iota=1$ surface. From Fig. (\ref{fig:gnet_neutral}),
the gradient in fast ion pressure is slightly larger in the region
$\rho=0.6-0.85$ with charge-exchange present, i.e. more like the
H-rich case than the He-rich case. But, there is in-fact a decreased
pressure gradient region in the peripheral region, i.e. $\rho>0.85$,
as indicated by the shading in the figure. Thus it is difficult to
make a clear conclusion from this analysis. A full simulation of the
fast ion pressure distribution in the presence of Helium background
plasma, might further alter the spatial gradient beyond the ``excluding-CX''
case: for example, as charge-exchange losses are still prevalent at
the edge of Helium discharges, they might tend to enhance the edge
fast-ion pressure gradient. Further modelling is required to solidify
this argument.

Mode damping may play a role in the apparent change of mode stability
going from H to He. Possible mechanisms include continuum damping,
mode coupling, electron collisional damping and Landau damping on
ions and electrons. Of all these mechanisms, we draw attention to
ones which would exhibit a change in behaviour going from H to He.
The obvious candidate is Landau damping on the ion distribution function.

For Landau damping, resonance with bulk plasma ions with the mode
has a stabilising effect, depending on the velocity gradient of the
ion distribution function $df/dv$ according to the terms in the Vlasov
equation. Landau damping is often determined by a resonance between
the mode and particular particles in the bulk distribution function,
where the magnitude of the interaction is determined by the ratio
of the phase velocity to the thermal velocity. For high frequency
fast ion instabilities like TAEs, this is often considered negligible,
because the mode velocity is much faster than the ion thermal velocity.
However, for the modes present here their low frequency means that
the phase velocity is much lower, so it is worthwhile considering
that Landau damping could play a role. For both the EIC mode at 8kHz
with m=1,n=1 and the mode at 30kHz with m=4,n=1 the phase velocity
(in the helical direction of propagation) is $v_{phase}\sim28$km/s.
This is considerably smaller than $v_{th}$, which for H discharges
at 1keV (near the $\iota=1$ surface) is $\sim400$km/s and $\sim200$km/s
for He. 

A precise evaluation of the role of Landau damping requires a model
for the specific type of MHD mode. One mode of interest to consider
is the ion acoustic oscillation \cite{landau}, which has a dispersion
relation given by $\omega/k=(1.45-0.6i)v_{th}$, where the negative
imaginary component represents Landau damping. Whilst the modes of
interest have much lower phase velocity than the thermal velocity,
they are in the acoustic range and demonstrate that Landau damping
could occur in this frequency range. Changing from He to H, the thermal
velocity changes a factor of 2 as it scales as $v_{th}\sim1/\sqrt{A}$.
If we consider the $v_{th}$ and charge (Z, controlling ion density
scaling for constant electron density) dependence of the ion distribution
function, i.e. $f(v)\sim Z^{-1}v_{th}^{-1}\exp(-v^{2}/v_{th}^{2})$,
we can estimate simply how the velocity space gradient $\partial f(v)/\partial v$
at $v=v_{phase}=28km/s$ changes from H to He. This gradient is $\approx4$
times larger in He than H, and so will contribute to greater Landau
damping in Helium. 

On the other hand, we note another damping effect: the stabilisation
of the EIC that can occur with ECH which increases temperature and
thus decreases the magnetic Reynolds number, which has the effect
or reducing the radial width of the interchange mode eigenfunction
\cite{du_ech_suppression_prl}. In this case, moving from H to He
would have the effect of reducing the magnetic Reynolds number, enlargening
the eigenfunction, so would not be the mechanism behind the stabilization.

\subsection{Role of EIC events on ion heating\label{subsec:Role-of-EIC}}

\begin{figure}
\includegraphics{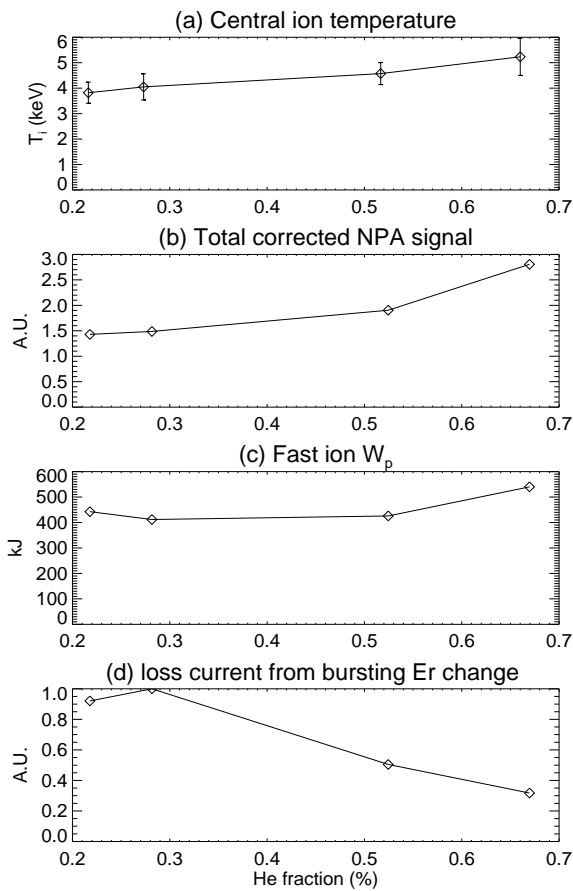}

\caption{Comparison of the He fraction dependence of: (a) Central ion temperature
averaged over $\rho<0.3$, (b) Integral of the corrected NPA signal
over energy, (c) Fast ion $W_{p}$ value obtained from analysis of
the diamagnetic loop and bulk contribution and (d) fast ion losses
calculated from PCI phase velocity using Eq. (\ref{eq:jr}).\label{fig:Comparison-of-analysis}}
\end{figure}

The key results from this paper are summarized in Fig. (\ref{fig:Comparison-of-analysis}),
including the He concentration dependence of (a) central ion temperature,
(b) the integral of the corrected NPA signal over energy, (c) the
fast ion $W_{p}$ component obtained from Sec. (\ref{subsec:Diamagnetic-stored-energy})
and (d) a rough estimate of the fast ion loss current $j_{r}$ from
the EIC events, obtained from PCI data, computed using Eq. (\ref{eq:jr}).
For increasing He concentration, the ion temperature increases and
the EIC burst rate goes down. Measures of confined fast ion density
increase.

From this correlation we can speculate whether the decreasing burst
rate / fast ion losses in He might be a cause for increasing ion temperature.
Whilst the improvement of transport is definitely expected from the
gyro-Bohm scaling, we note from Sec (II) that the improvement was
not enough to fully explain the temperature increase particularly
near the edge of the plasma. The increase in the fast ion energy content
in Fig. (\ref{fig:Comparison-of-analysis}b,c) in He discharges might
further explain the strong ion temperature dependence. In particular,
it was shown that the edge pedestal value of ion temperature increased
with He concentration \cite{tanaka_nf2017}. This increase may be
due to a reduction of fast ion losses, or an increase in the direct
charge-exchange losses of perpendicularly injected fast ions to background
neutrals, which are more abundant in Hydrogen, or a combination of
both. 

The 2D PCI diagnostic clearly revealed bursting events where the phase
velocity reversed to the electron diamagnetic direction in the region
$\rho=0.6-0.9$, and is attributed to the change of electric field
associated with fast ion losses on the basis with correlation with
fast ion diagnostics in Sec. (\ref{subsec:Fast-ion-loss}) including
a lost ion RF spectrometer, the compact NPA system, as well as the
diamagnetic loop signal. It is difficult to assess the magnitude of
losses arising from the fast-ion MHD with present diagnostics. On
one hand, the changes in the diamagnetic stored energy with each burst
are only around 1kJ out of a total of 1MJ, which would suggest that
the losses are not significant. However, the redistribution of fast
ions may mask any local change of the fast ion pressure. Furthermore,
it is not clear, from the correlation analysis in Fig. (\ref{fig:cc_anderes}e),
that the stored energy is saturating in-between burst events, suggesting
that the $\Delta W_{p}$ value for each burst is not indicative of
the net effect for a series of MHD events. 

\section{conclusions\label{sec:Discussion-and-conclusions}}

\emph{Observation of fast ion MHD: }This activity consists of EICs
at around 8kHz with m/n=1/1 and an unidentified mode at 30kHz with
m/n=4/1. These EIC bursts consist of marked changes in the radial
electric field, which is derived from the phase velocity of turbulence
measured with the 2D phase contrast imaging (PCI) system. Similar
bursts are detected in edge fast ion diagnostics including CNPA, SiFNA
and ICE RF spectrometer channels at 300,400MHz. These EIC burst events
appear to be similar in character though weaker than those reported
in \cite{eic_nf}.

\emph{Stablity of EIC affected by ion species:} The change of the
nature of the bursts with He concentration may help to understand
the destabilisation mechanism of EIC bursts. Change of drive and damping
mechanisms related to the change of mass and charge of Helium may
be playing a role, including change of the fast ion pressure gradient
near the 1/1 rational surface, and damping of the mode on the bulk
ion distribution function. In Helium, there is a longer time between
events as shown by Fig. (\ref{fig:br_vr_mag}a), and EIC amplitude
is larger as shown in Fig. (\ref{fig:br_vr_mag}d). However, they
result in very similar fast particle losses characterized by the change
in electric field indicated in Fig. (\ref{fig:br_vr_mag}b). This
behaviour is similar to that recently shown by Deuterium discharges
\cite{d_ohdachi}.

\emph{Effect of EIC-induced fast-ion losses on ion heating power:}
Whilst the fast ion diagnostics are not quantitative enough to clarify
the magnitude of the fast-ion losses from each event, the observed
correlation of the fast ion loss current with with the confined fast-ions
and the ion temperature suggests that there may be a significant amount
of power lost from each EIC event. This was suggested in the original
article on the EIC \cite{eic_nf}.

\emph{Turbulence suppression:} Fluctuation amplitude is suppressed
during the EIC bursts. Although the average turbulence level stays
independent of Helium fraction, there is lower edge turbulence in
the phases of the MHD bursts, which are more prevalent in H-rich discharges.
However, this is offset by a larger level of turbulence in the period
between these bursts, most particularly in H rich discharges. Thus,
the result is consistent with gyro-Bohm scaling.

The implication of this work is that fast ion instabilities must be
considered in any transport, or turbulence study, just as with bulk
plasma instabilities. Energy lost from fast ion instabilities may
degrade the heating power and thus affect any energy transport analysis
carried out, and the changes in the radial electric field, time averaged,
can considerably change the transport properties. Also, change in
working gas can dramatically affect fast ion instabilities, either
through change of drive, or through Landau damping on bulk ions. This
is analogous, but different to the stabilization of the EIC that can
occur due to the Magnetic Reynolds number changing at higher temperature
\cite{du_ech_suppression_prl}.

Now that the first LHD Deuterium campaign has concluded these results
will complement the understanding of a comparison of EIC events between
Hydrogen and Deuterium (D) plasma. Bursts in D plasmas with D perpendicular
beams are larger and less frequent than those of H beams into an H
plasma. This is similar to the results reported in this paper where
H beams injected into He plasma are larger and less frequent than
Hydrogen plasma. Furthermore, the EIC bursts in D plasma can induce
a significant loss a of neutron yield \cite{d_ohdachi}, with a long
time to recover between events, suggesting that the EIC can cause
a large loss of heating power.

Further analysis is required to understand the drive of the 30kHz
mode observed and to connect that to the fast ion instabilities. Since
this mode bursts in amplitude with the EIC events, and has the same
poloidal propagation velocity it suggests that it may be a coupling
with the EIC event and/or with the precession of trapped fast ions
from perpendicular beam injection. It is also a mystery of why the
8kHz EIC and the 30kHz mode propagate in the electron diamagnetic
direction despite the deeply trapped beam ions propagating in the
ion diamagnetic direction. Improvements in modelling of fast ions
and background neutrals to include bulk Helium in the AURORA and GNET
codes would enable understanding of the mode suppression in Helium.
Finally, the hypothesis of the role of Landau damping in on the EIC
would be solidified with a more detailed analysis of the wave/particle
interaction.

\section*{Acknowlegements}

C. Michael gratefully acknowledges the support of the NIFS directorate
for the invitation as a visiting associate professor for the months
of June-July 2016. He also acknowledges gratefully colleagues at NIFS
for being responsive to providing data and interpretation. We would
like to acknowledge K. Toi, Professor Emeritus of NIFS and X. Du of
UC Irvine for invaluable discussions about this topic. Work on this
paper was also funded by the Australian Government's National Collaborative
Research Infrastructure Strategy (NCRIS), NIFS grants NIFSULHH013,
NIFS10ULRR702, NIFSULHH004, NFSULHH005, NIFSULHH02, NIFSULHH028, NIFS16KLER045,
NIFS14UNTT006 and JSPS kakenhi grant number 16H04620p. \bibliographystyle{aip}
\bibliography{lhdpaper_library}

\end{document}